\documentstyle[11pt,paspconf,epsf]{article}
\begin{document}

\title{The H\,I fine structure of HVC\,187 near NGC\,3783:\\
gas in the leading bridge of the Magellanic System}

\author{Bart P. Wakker, Blair D. Savage}
\affil{Department of Astronomy, University of Wisconsin, Madison, WI 53706}
\author{Tom A. Oosterloo}
\affil{CNR, Istituto di Fisica Cosmica, Milan, Italy}
\author{Mary Putman}
\affil{Mount Stromlo and Siding Springs Observatories, Australia}


\def\HI{H{\small I}}

\begin{abstract}
We present an analysis of high-resolution \HI\ data of one of the cores of
HVC\,187 (Wakker \& van Woerden 1991), HVC287+22+240. Structure is present down
to the lowest-measurable scale ($\sim$1 arcmin) and several concentrations
appear to be unresolved. Most of the cores seen at low resolution break up into
smaller subcores at higher resolution. The typical volume density and pressure
are estimated to be $\sim$30\,R$^{-1}$\,D$^{-1}_{\rm kpc}$\,cm$^{-3}$ and
$\sim$18000\,R$^{-1}$\,D$^{-1}_{\rm kpc}$\,K\,cm$^{-3}$, respectively, where R
is the resolution in arcmin and D$_{\rm kpc}$ the unknown distance in kpc.
\end{abstract}

\section{Background}
Using the Goddard High Resolution Spectrograph (GHRS) on the Hubble Space
Telescope (HST), Lu et al.\ (1998) measured the column density of S\,{\small II}
toward HVC\,187, using the Seyfert galaxy NGC\,3783 as a background probe. This
object is one of the main extreme-positive-velocity clouds, defined by Wakker \&
van Woerden (1991). Using the \HI\ column density derived from the data
discussed here, the sulphur abundance was found to be 0.25$\pm$0.07 solar,
similar to the value in the Magellanic Clouds. Combining this with the tidal
model of Gardiner \& Noguchi (1996), Lu et al.\ concluded that the HVC is part
of the leading arm of the Magellanic System, implying a distance of 10--50\,kpc.
\par Here we present the \HI\ data used to derive the \HI\ column density.
Observations with the Australia Telescope Compact Array (ATCA) were done in June
1994 and March/June 1995, with a resolution of 60\arcsec x38\arcsec. The primary
beam had an FWHM of 33\farcm7. For calibration, mapping and analysis the MIRIAD
software package was used. Cleaning was done using the Multi-Resolution Clean
(Wakker \& Schwarz 1987). The ATCA data were combined with data from the Parkes
Multibeam survey (Staveley-Smith 1997). Figure 1 shows the Parkes map of
HVC\,187. Other HVCs, with velocities of about +100\,km/s, are present in this
same region (complex~WD).

\begin{figure} 
\plotfiddle{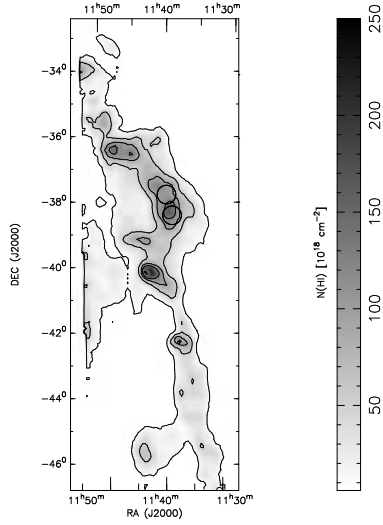}{6cm}{0}{55}{55}{-150}{-30}
\caption{
\HI\ column density map of HVC\,187 between +204 and +270\,km/s, based on Parkes
data (16\farcm7 resolution). Contours are at column densities of 10, 50, 90 and
130$\times$10$^{18}$\,cm$^{-2}$. The circles near dec.\ $-$38\deg\ show the HPBW
of the interferometer fields.}
\end{figure}

\begin{figure} 
\plotfiddle{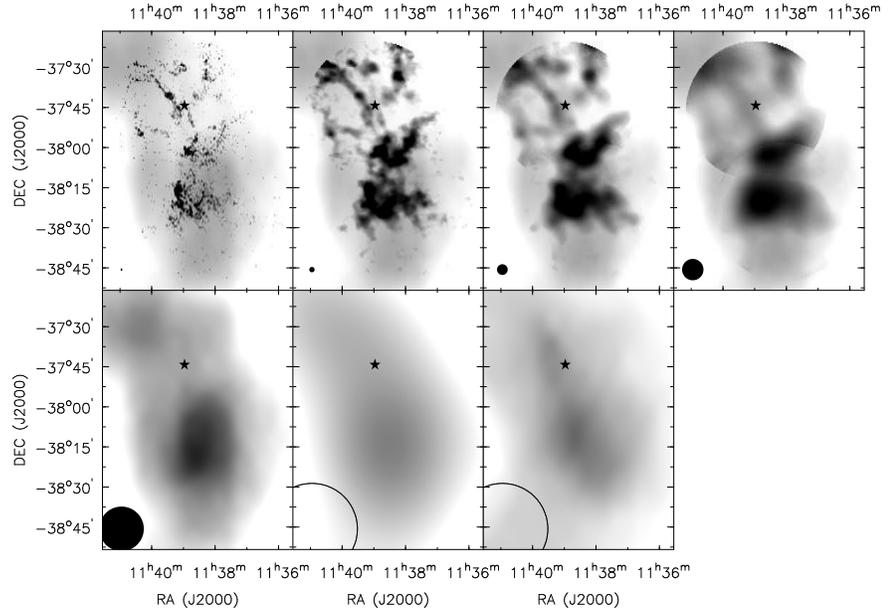}{6.5cm}{0}{80}{80}{-190}{-60}
\caption{
\HI\ column density maps (combined ATCA and Parkes data). Top row, left to
right: maps at original 60\arcsec x38\arcsec\ resolution, and at 2\arcmin
x2\arcmin, 4\arcmin x4\arcmin\ and 8\arcmin x8\arcmin. Bottom row, left to
right: Parkes map (16\farcm7 beam), Parkes map smoothed to 34\arcmin, and Villa
Elisa map (34\arcmin\ beam; Morras \& Bajaja 1983). The FWHM sizes of the beams
are indicated in the lower left corner of each panel. The black star shows the
position of NGC\,3783.}
\end{figure}

\section{\HI\ maps}
\par Figure 2 shows the total column density map of HVC287+22+240, based on
combining ATCA and Parkes data. These maps were created by first summing all
interferometer channels, applying a mask. The mask was created by using a
3-sigma cutoff in a map smoothed to double the beam, so that pixels with low
S/N ratio could be eliminated. The summed maps were then corrected for primary
beam attenuation, converted to brightness temperature, and integrated over
velocity. Finally, the two fields were combined. Next, the interferometer and
single-dish maps were combined following the method described by Schwarz \&
Wakker (1991). This consists of smoothing the primary-beam corrected
interferometer map to the single-dish beam and subtracting the result from the
single-dish map to find the emission that was filtered out by the lack of short
interferometer baselines. Adding back the cleaned interferometer map gives the
final map.
\par The resulting map shows the presence of structure at scales of 1 arcmin.
Smoothing decreases the apparent column density in the concentrations,
indicating that they are not fully resolved. As is usual for high-velocity cloud
cores the simple structure seen at $>$20\arcmin\ resolution resolves into many
scattered cores.

\section{Spin temperature limits}
\par Figure 3 shows spectra toward the six brightest continuum sources in the
field. These spectra were made using the method described by Wakker et al.\
(1991), using only baselines longer than 180\,m. The profiles were extracted,
converted to brightness temperature, corrected for primary beam attenuation, and
hanning smoothed over 5 points. Table~1 collects the measurements. The spin
temperature $T_s$ follows from the \HI\ emission ($T_e$) and the optical depth
of the absorption ($\tau$): $T_e = T_s (1 - e^{-\tau})$. We included in $T_e$ a
$\sim$2\,K contribution from a smooth background (filtered out by the
interferometer).

\begin{table}[ht]
\begin{tabular}{rrrrrrrrr}
\noalign{Table 1}\noalign{\smallskip}\hline\noalign{\smallskip}
\# &   RA\ \ \ &     DEC\ \  & Flux\ \ \ \  &T$_{\rm B}$\ &$\tau_{\rm max}$
                                                         &S/N &T$_{\rm e}$
                                                                  &T$_{\rm s}$\\
   &h\ \ m\ \ s\ \ \ &d\ \ \arcmin\ \ \arcsec\ \ \ 
                             &  mJy\ \ \ \  & K\ \ &     &    &  K\ \  & K\ \ \\
\hline\noalign{\smallskip}
(1)& (2)\ \ \  &   (3)\ \ \  &  (4)\ \ \ \  &  (5) & (6) & (7)&  (8)   & (9)  \\
\noalign{\smallskip}\hline\noalign{\smallskip}
 1 &11 39 34.4 & $-$37 48 37 & 75.8$\pm$1.4 & 26.8 & 0.2 & 15 & $<$1.6 & $>$19\\
 2 &11 39 02.0 & $-$37 44 16 & 38.7$\pm$1.2 & 13.7 & 0.4 &  8 & $<$2.2 & $>$12\\
 3 &11 39 49.2 & $-$37 48 11 & 45.0$\pm$1.5 & 15.9 & 0.4 &  8 & $<$1.8 & $>$10\\
 4 &11 37 34.3 & $-$37 49 55 & 45.5$\pm$2.7 & 16.1 & 0.9 &  4 &   =3.8 & $>$ 9\\
 5 &11 38 59.3 & $-$38 00 47 & 33.2$\pm$2.1 & 11.7 & 1.1 &  4 &   =7.4 & $>$14\\
 6 &11 40 32.4 & $-$37 40 47 & 43.1$\pm$2.9 & 15.2 & 1.2 &  4 & $<$3.4 & $>$ 8\\
\noalign{\smallskip}\hline\noalign{\smallskip}
\end{tabular}
Col.~1: identification; Col.~2 and 3: position; Cols.~4 and 5: 21-cm continuum
flux and corresponding brightness temperature; Col.~6: upper limits to optical
depth; Col.~7: S/N ratio of continuum spectrum; Col.~8: peak brightness
temperatures or limits on \HI\ emission near the continuum source. Col.~9: lower
limits on the spin temperature.
\end{table}

\begin{figure} 
\plotfiddle{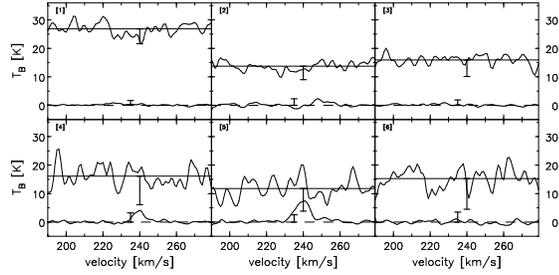}{2.9cm}{0}{58}{58}{-170}{-45}
\caption{
Top lines: emission from the continuum source. Bottom lines: \HI\ emission
spectra, found by averaging in a 9x9 pixel box centered on the continuum source,
subtracting out the central 5x5 pixels. The vertical bars indicate 3$\sigma$
absorption or emission.
}\end{figure}

\section{Linewidths, densities and pressures}
\par Several concentrations can be discerned (see Figure 2). Figure~4 shows that
the brightness temperature decreases with resolution in a manner similar to the
expected relation for a 1\arcmin--2\arcmin\ gaussian, i.e., they are (almost)
unresolved. The run of column density with distance to the center of the blob
shows the presence of column density contrasts of a factor 3 on arcminute
scales.
\par Gaussian fits were made to the spectrum at each pixel. Linewidths are
5--10\,km/s, increasing slightly with resolution due to beam-smearing of
velocity gradients. The components with more extreme velocities tend to occur
away from the brightest core at 11$^{\rm h}$39$^{\rm m}$, $-$38\deg15\arcmin. In
low resolution single-dish spectra this would show up as a broadening of the
line profiles away from the core. Assuming that the line-of-sight density
profile is gaussian, combining the angular diameter ($\alpha$), distance (D) and
column density (N$_{\rm H}$) of a concentration yields the volume density: $n =
{\rm N}_{\rm H} /(1.064 \alpha$ D). The linewidth, $W$ (in km/s), is converted
to a temperature: $T = 21.8 W^2$. The pressure then is $P = nT$. The derived
apparent densities and pressures tend to depend on resolution ($R$), because the
measured radius is proportional to $\sqrt{R}$, and the column density to
1/$\sqrt{R}$. The average measured densities of the cores are of the order of
30\,R$^{-1}$\,D$^{-1}_{\rm kpc}$\,cm$^{-3}$, pressures are
18000\,R$^{-1}$\,D$^{-1}_{\rm kpc}$\,K\,cm$^{-3}$. These pressures represent the
combined thermal and ``turbulent'' pressure of the (presumably) cool \HI\ cores.

\begin{figure}
\plotfiddle{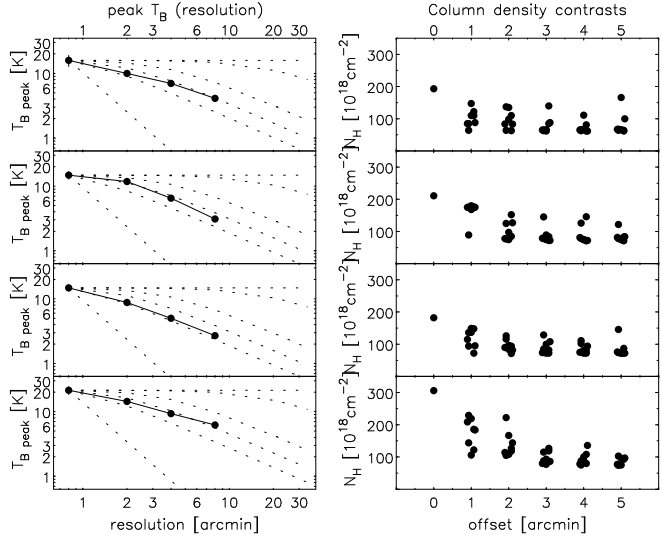}{3.8cm}{0}{90}{90}{-330}{-90}
\caption{
Measurements on concentrations. Left: brightness temperature as function of
resolution. Dotted curves give the relations for a point source, a 1\arcmin,
2\arcmin, 4\arcmin, 20\arcmin\ gaussian, and an extended source. Right: run of
column density with distance to the center of the concentration.}
\end{figure}

\begin{figure}
\plotfiddle{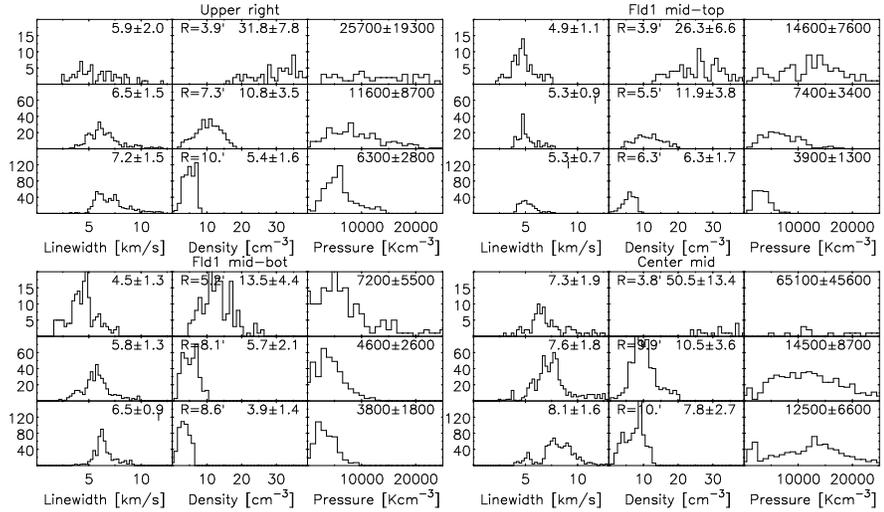}{1.9cm}{0}{80}{80}{-230}{-360}
\caption{
Linewidth, volume density and pressure distribution for all pixels in a
concentration for which the column density is $>$0.5 times the peak value. Nine
plots are shown for each concentration, in three rows of three, each row
corresponding to 1\arcmin, 2\arcmin\ or 4\arcmin\ resolution.}
\end{figure}

\end{document}